\documentclass[aip,jap,reprint]{revtex4-1}

\draft % marks overfull lines with a black rule on the right

\usepackage{graphicx}% Include figure files
\usepackage{dcolumn}% Align table columns on decimal point
\usepackage{bm}% bold math
\usepackage{color}% bold math

\begin{document}

\title{Multifrequency transverse Faraday effect in single magneto-dielectric microspheres} 
\affiliation{School of Physics, The University of Western Australia, 35 Stirling Highway, Crawley WA 6009, Australia}
\author{Ivan S. Maksymov}
\email[Ivan S. Maksymov ]{ivan.maksymov@uwa.edu.au}
\author{M. Kostylev}

\date{\today}

\begin{abstract}

We propose using a single magneto-dielectric microsphere as a device for enhancing the transverse Faraday effect at multiple wavelengths at the same time. Although the diameter of the sphere can be $<1$ $\mu$m, the numerically predicted strength of its magneto-optical (MO) response can be an order of magnitude stronger than in MO devices based on thick magnetic plates. The MO response of a microsphere is also comparable with that of subwavelength magneto-dielectric gratings which, however, operate at a single wavelength and occupy a large area. In contrast to gratings and thick plates, the compact size of the microsphere and its capability to support spin-wave excitations make it suitable for applications in nanophotonics, imaging systems, and magnonics.

\end{abstract}

\maketitle %\maketitle must follow title, authors, abstract and \pacs

\section{Introduction}

Magneto-optical (MO) effects have found important applications in data storage, telecommunications, imaging, spectroscopy,\cite{1,2,3,4} as well as in magnonics. \cite{5,6} An increasingly broad application of nanotechnologies in these areas sets new requirements for novel MO devices by demanding, among other things, a stronger MO response and smaller dimensions.

For example, small dimensions are important for $3$D imaging systems based on holographic principles.\cite{4} Such systems require a spatial light modulator (SLM) -- a device used to modulate amplitude, phase, or polarisation of light waves in space and time. Basically, an SLM device can produce high quality $3$D images if it possesses pixels sizes $\leq 1$ $\mu$m and also is fast enough to address a large number of pixels within a single image frame. SLM devices exploiting MO effects have a very fast response time. However, their typical pixel sizes are $\sim 10$ $\mu$m.\cite{4}

At the nanoscale, MO effects can be enhanced by using subwavelength diffraction gratings \cite{7,8,9,10,11}, magneto-photonic crystals, \cite{12} and nanoantennas, \cite{13,14} which are made of pure magnetic materials or consist of alternating magnetic-nonmagnetic layers. However, as a typical grating period is $\sim500$ nm and there should be tens of periods to achieve optimal optical properties, the footprint of the grating is large. Magneto-photonic crystals often have comparable dimensions and suffer from similar disadvantages.  

Typical dimensions of ferromagnetic/normal metal nanoantennas \cite{13,14,15,16} can be much smaller than $1$ $\mu$m because of magneto-plasmonic resonances which give an additional degree of freedom for light manipulation at the nanoscale.\cite{15} However, due to huge absorption losses in ferromagnetic metals, plasmonic resonance properties of such nanoantennas are not very strong as compared with nanoantennas fabricated of gold or silver (nonmagnetic metals with relatively low absorption losses). Consequently, measures have to be taken to mitigate absorption losses, which can be done by reducing the amount of ferromagnetic metals and tailoring Fano resonance effects.\cite{16}. Alternatively, by analogy with subwavelength gratings,\cite{11} losses in nanoantennas can be reduced by using magneto-dielectrics instead of ferromagnetic metals.   

Normally, magnetic gratings, photonic crystals and nanoantennas are used to enhance the Faraday rotation and Kerr effect in different configurations. \cite{15,17} For instance, if the incident light is $p$-polarised and the static external magnetic field is applied perpendicularly to the plane of incidence, the so-called transverse MO Kerr effect (TMOKE) \cite{1,2,3,15} is observed in the reflection mode.

Less well-known are transverse MO effects observed in the transmission mode. These effects are similar to the TMOKE but occur in transparent magnetic films or plates \cite{2,15,18}. An example of a transverse MO effect in the transmission mode is the transverse Faraday effect (TFE). \cite{18} This effect can be used in imaging, data storage systems and magnonics.

The TFE has been observed in transparent magnetic plates and Fabry-Perot resonators. \cite{2,18} Because the TFE response is amplified as a result of the wave propagating back and forth within a magnetised medium, it is enhanced in a Fabry-Perot resonator. However, a typical Fabry-Perot resonator consists of a $20$ $\mu$m-thick plate of a transparent magnetic material sandwiched between a pair of mirrors, which in turn consist of $10$ alternating layers of high- and low-refractive-index quarter-wave-thick dielectrics.\cite{18} Consequently, the dimensions of the resonator are also very large, which makes it unsuitable for miniaturisation.

\begin{figure*}[htb]
\centering\includegraphics[width=17cm]{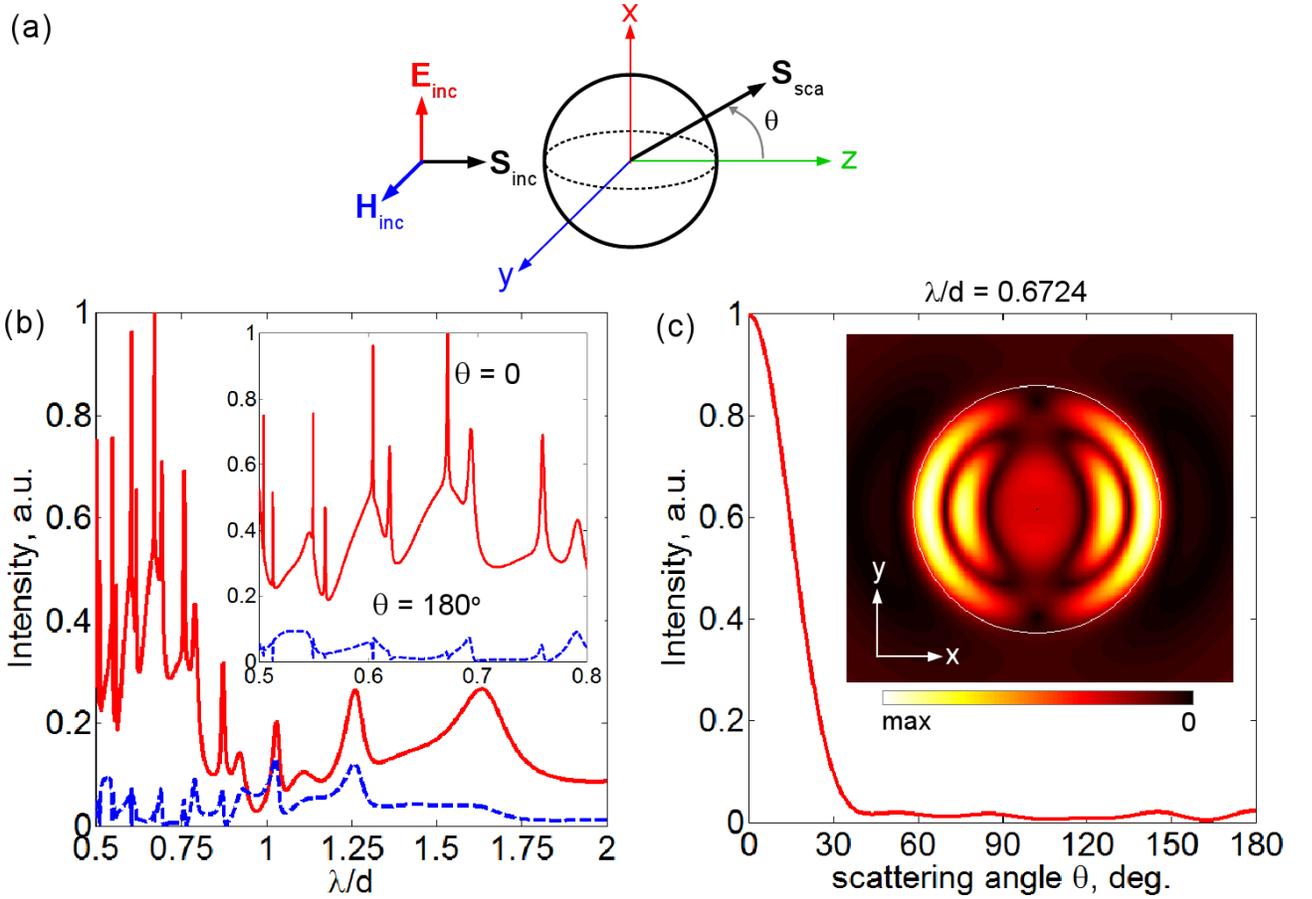}
\caption{(a) Schematic of the scattering by a sphere. The incident plane wave propagates along the \textit{z}-axis, the optical electric and magnetic fields are polarised along the \textit{x} and \textit{y} axis, correspondingly. Results obtained using the Mie theory for nonmagnetic spheres: (b) The intensity of light scattered by the homogeneous nonmagnetic sphere in the forward (solid line) and backward (dashed line). (c) Far-field emission pattern and the modulus of the Poynting vector at $\lambda / d=0.6724$, which corresponds to the maximum of the forward-scattered intensity.} 
\end{figure*}

In this paper, we predict and demonstrate theoretically a large multifrequency TFE in single microspheres made of Bi$^{\rm{3+}}$-substituted yttrium-iron magnetic garnet (Bi:YIG). Bi:YIG is a magneto-dielectric exhibiting large MO activity and high transparency in the visible and infrared spectral ranges.\cite{3} By exploiting high-quality factor resonances of a Bi:YIG microsphere we show a strong TFE response at multiple wavelengths at the same time. The strength of the predicted TFE is comparable with the strength of the TMOKE in subwavelength magnetic gratings. \cite{9,11} Moreover, the observed TFE response is higher than that in micron-thick transparent magnetic plates. \cite{18} However, the dimensions of a single sphere are significantly smaller as compared with the area occupied by a subwavelength grating or the thickness of a plate used in modern MO devices.

\section{Results and discussion}

The problem of light scattering by a homogeneous sphere of arbitrary diameter and dielectric permittivity is exactly soluble by using the Mie theory.\cite{19,20} Whereas the Mie theory can be extended to calculate the scattering by homogeneous magnetic spheres described by their magnetic permeability \cite{20}, the problem of light scattering by a sphere magnetised along a certain coordinate direction is more difficult. This is because the dielectric permittivity becomes a tensor describing the interaction between the light and the static external magnetic field (or the internal magnetisation of the medium)\cite{2}

\begin{eqnarray}
\epsilon = 
\left(
\begin{array}{cccc}
\epsilon_{\rm{xx}}&\epsilon_{\rm{xy}}&\epsilon_{\rm{xz}} \\
\epsilon_{\rm{yx}}&\epsilon_{\rm{yy}}&\epsilon_{\rm{yz}} \\
\epsilon_{\rm{zx}}&\epsilon_{\rm{zy}}&\epsilon_{\rm{zz}}
\end{array}\right)
\label{eq:one}.
\end{eqnarray}

\noindent By considering Bi:YIG as an isotropic material, the three diagonal elements of $\epsilon$ become identical, and in the presence of a static external magnetic field along the \textit{y}-axis, there is a non-zero off-diagonal element $\epsilon'$, which couples the \textit{x}- and \textit{z}-components of the optical electric field

\begin{eqnarray}
\epsilon = 
\left(
\begin{array}{ccc}
n^2&0&\epsilon ' \\
0&n^2&0 \\
-\epsilon '&0&n^2
\end{array}\right)
\label{eq:two}.
\end{eqnarray}

Due to low absorption losses in Bi:YIG the refractive index $n$ and $\epsilon'$ of Bi:YIG can be assumed to be real and also frequency-independent over a narrow range of wavelengths: $n=2.2$ and $\epsilon'={\rm{i}}0.005$. \cite{3,21} Because Bi:YIG is a ferrimagnetic material, the spin-orbit coupling is the dominant source of the MO interaction and it makes $\epsilon'$ proportional to the magnetisation $M$ of the medium. \cite{1,2,3}

\begin{figure*}[htb]
\centering\includegraphics[width=17cm]{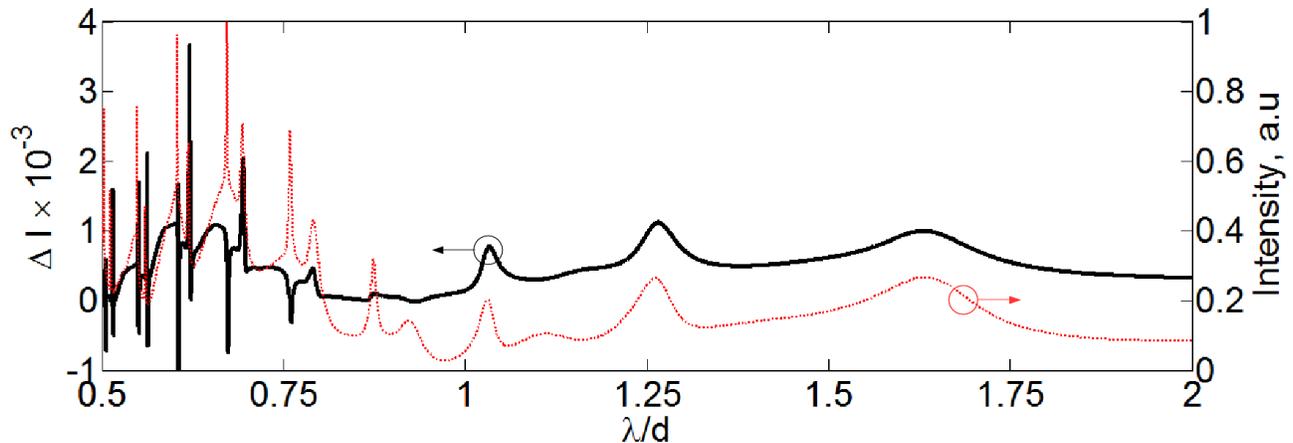}
\caption{Result obtained using the FDTD method: Solid line -- the forward-scattering intensity differential $\Delta I$ quantifying the strength of the TFE in the microsphere as a function of the normalised wavelength $\lambda/d$. The intensity of the forward-scattered light (dashed line) is also plotted to show the direct correlation between the maxima of $\Delta I$ and the resonances of the microsphere. The microsphere is uniformly magnetised along the \textit{y}-axis in Fig. $1$(a). } 
\end{figure*}

By considering the off-diagonal elements of $\epsilon$ in the Mie theory, one can take into account the physical mechanisms responsible for the MO Kerr and Faraday effects contribution to the optical response of a sphere. \cite{22, 23} It has been shown that the MO Kerr effect in a homogeneous cobalt sphere is small but detectable.\cite{23} The small MO response of cobalt spheres is due to large optical\cite{23_1} and magneto-optical\cite{23_2} absorption effects in this ferromagnetic metal. Considerable optical absorption losses are typical of all metals, including gold and silver. These losses have long been known to be the drawback of metallic nanostructures used in photonic devices such as, e.g., nanoantennas. \cite{24} Consequently, a large and growing body of research investigates all-dielectric nanostructures, e.g., low-loss single-sphere nanoantennas. \cite{24,25} In this context, the use of Bi:YIG as the model material of the microsphere opens up opportunities to overcome the drawback of metallic ferromagnetic spheres. 

It is also worth noting Brillouin light scattering (BLS) from spin-wave modes in uniformly magnetised ferromagnetic and ferrimagnetic spheres. \cite{26,27,28} The fluctuation of magnetisation due to spin-wave modes causes a time-dependent change of the dielectric permittivity tensor $\delta \epsilon$ of the material of the sphere. The contribution of the transverse magnetisation to $\delta \epsilon$ allows generating a description of BLS by processes in which a single spin-wave quantum (magnon) is created or destroyed. In this case, Bi:YIG still can be used as the model material of the microsphere because BLS has been used to study nonlinear spin wave phenomena in Bi:YIG structures.\cite{29} Furthermore, microspheres can also be made of pure yttrium iron garnet (YIG), which is a very well-known material used in magnonics.\cite{6}

Finally, both semiconductor and ferromagnetic properties have been established in some of the rare earth mononitrides, which thus attract interest for the potential to exploit the spin of charge carriers in spintronics. \cite{30} The refractive index of these materials is $n \approx 2$ and they are also transparent in the visible and infrared spectral ranges.\cite{31} Most significantly, these materials also exhibit significant MO activity.\cite{30} Thus, they might be employed instead of Bi:YIG. 

Fig. $1$(a) schematically shows the scattering by an arbitrary homogeneous nonmagnetic sphere. By applying the Mie theory for nonmagnetic spheres \cite{19} we show that the intensity of the forward scattered light [$\theta = 0$ in Fig. $1$(a)] has multiple peaks in the spectral range $0.5-0.8 \lambda/ d$, where $d$ is the diameter of the sphere and $\lambda$ is the wavelength in the free space [Fig. $1$(b)]. The intensity of the backscattered light ($\theta = 180^{\rm{o}}$) in this spectral range is low. This implies that at the resonance the single microsphere has a very directive far-field emission pattern [Fig. $1$(c)].

For example, in order to achieve the maximum forward-scattering intensity in the visible spectral range a BI:YIG sphere with the diameter $d \leq 1$ $\mu$m is needed. It is noteworthy that a nonmagnetised sphere with $d \approx 1$ $\mu$m will be in the so-called multi-stripe domain state.\cite{32} Basically, light can be scattered by the stripe domain structure.\cite{33} However, the strength of this process is very low and it can be neglected, i.e. we will assume that a nonmagnetised Bi:YIG sphere has the optical properties of a homogeneous nonmagnetic sphere with the same refractive index.

The application of the static magnetic field orientated along the \textit{y}-axis [Fig. $1$a] will magnetically saturate the sphere by aligning the direction of magnetisation inside the sphere along the \textit{y}-axis. Therefore, in the saturated state the MO properties of the sphere can be modelled by using the tensor $\epsilon$ [Eq. ($1$)]. In this case, the conventional Mie theory cannot be applied; it must be extended employing, e.g., a perturbation approach.\cite{23} Alternatively, approximate approaches such as, e.g., a modified discrete dipole approximation (DDA) \cite{34,14} may be employed. However, these methods have disadvantages. The perturbation approach requires the application of the Green's function in spherical coordinates.\cite{23} The application of the DDA to large MO spheres may require unaffordable computational efforts.\cite{35}

Consequently, we will use a finite-difference time-domain (FDTD) method. \cite{35_1} Although the FDTD method is also computationally demanding because it uses the staircase approximation of the surface of the sphere and thus requires a very fine finite-difference grid, its application to large Mie scattering problems is known to produce accurate results.\cite{36} The computing power of a modern desktop computer is enough for these simulations. Furthermore, the FDTD can be applied to scatterers with a more complex shape than the sphere. Of course, in this case for scatterers of a complex shape one also needs to solve a micromagnetics problem to find the distribution of the magnetisation inside the scatterer.  

The solid line in Fig. $2$ shows the forward-scattering intensity differential $\Delta I = I(M_{\rm{s}})-I(-M_{\rm{s}})$ quantifying the strength of the TFE in the microsphere as a function of the normalised wavelength $\lambda/d$. In this case, $I$ is the intensity of the forward-scattered light (shown for reference in Fig. $2$ by the dashed line) and $M_{\rm{s}}$ is the saturation magnetisation for Bi:YIG. The change in the sign of $M_{\rm{s}}$ implies the change in the direction of the static external magnetic field by $180^{\rm{o}}$ and it leads to the change in the sign of $\epsilon'$.\cite{2}

Recall that a similar differential $\Delta R = R(M_{\rm{s}})-R(-M_{\rm{s}})$, being $R$ the reflectivity, is used to quantify the strength of the TMOKE in gratings and other MO devices operating in the reflection mode.\cite{15,9,11} However, in contrast to all-magneto-dielectric gratings operating at a single wavelength,\cite{11} in Fig. $2$ the multiple maxima of $\Delta I$ are observed at a multiple wavelength corresponding to the sharp resonances supported by the microsphere. We note that the peak values of $\Delta I$ are of the same order of magnitude as the value of $\Delta R$ in the gratings.\cite{11} But the footprint of a single microsphere is much smaller than the area occupied by a grating.

\begin{figure}[htb]
\centering\includegraphics[width=8.5cm]{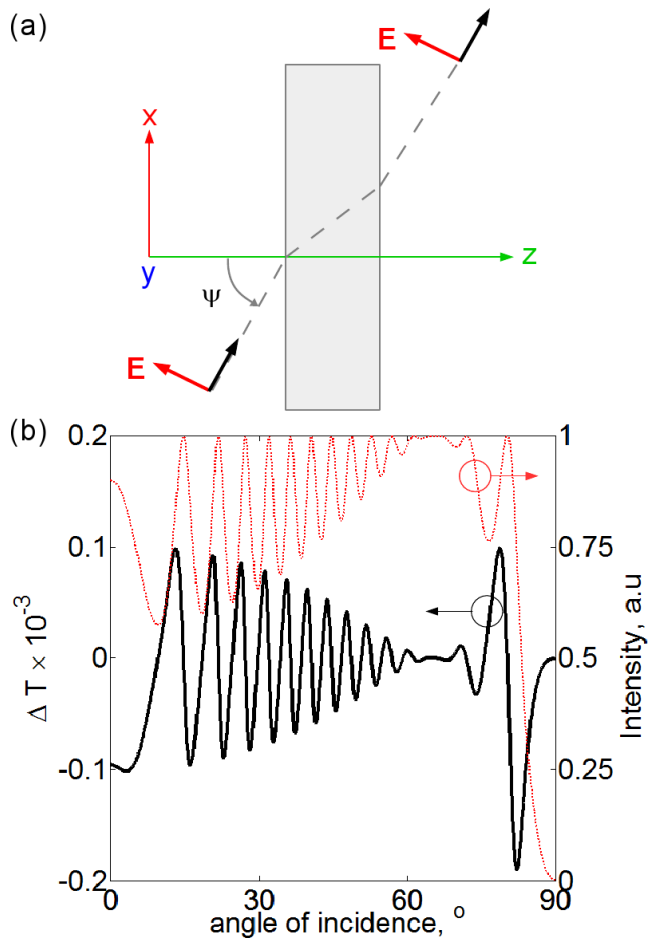}
\caption{ (a) Schematic of the transmission of the \textit{p}-polarised light through a plate uniformly magnetised along the \textit{y}-axis. The plate is assumed to be infinite along the \textit{x}- and \textit{y}-axis. (b) Solid line -- the transmission differential $\Delta T$ quantifying the strength of the TFE in the plate as a function of the angle of incidence $\psi$ for the wavelength in the free space $\lambda = 672.4$ nm. Dashed line -- the transmittance $T(0)$ of the nonmagnetised plate. Thanks to the one-dimensional character of the problem, these results were obtained using the Fresnel formula for \textit{p}-polarised light. \cite{2}} 
\end{figure}

To further demonstrate the advantages of the microsphere over the existing MO devices used to enhance transverse MO effects, we calculate the TFE in a $20$ $\mu$m-thick plate made of Bi:YIG [Fig. $3$(a)]. Such plates have practical applications in Fabry-Perot resonator-based MO devices.\cite{18} The considered plate is a one-dimensional scatterer with the spatial variation along the \textit{z}-direction so that $\frac{\partial}{\partial x} = \frac{\partial}{\partial y} = 0$. Therefore, for a \textit{x}- or \textit{y}-polarised incident wave, there is no \textit{z}-component of the optical electric and magnetic fields, and for the incident \textit{p}-polarised light the medium of the plate exhibits an effective refractive index $n_{\rm{eff}} = [n^2 + (\epsilon'^2/n^2)]^{1/2}$ (the medium does not exhibit MO activity for the \textit{s}-polarised light).\cite{2,18} Due to the dependence of $n_{\rm{eff}}$ on $\epsilon'$ the magnitude of the transmitted \textit{p}-light is sensitive to the magnetisation. Importantly, the TFE is not bipolar, so changing the direction of the static external magnetic field from $+y$ to $-y$ (i.e. changing $\epsilon' \rightarrow -\epsilon'$) does not alter the magnitude of the transmission $T$. Hence, the TFE is quantified by the differential $\Delta T = T(0)-T(M_{\rm{s}})$.\cite{2,18}

Importantly, the TFE can take place only in scatterers where a nonhomogeneous or leaky refracted wave is induced by a wave obliquely incident at the interface, i.e. the TFE vanishes at both the normal and grazing incidence.\cite{37} Consequently, in Fig. $3$(b) we plot $\Delta T$ as a function of the angle of incidence $\psi$ for the wavelength in the free space $\lambda = 672.4$ nm. We note that at this wavelength a sphere with $d=1$ $\mu$m would produce the maximum of the TFE response (Fig. $2$). Figure $3$(b) shows that in the absence of the static external magnetic field the transmission curve $T(0)$ for the \textit{p}-polarised light has multiple maxima due to interference effects. The curve of the differential $\Delta T$ follows the behaviour of $T(0)$. Around the Brewster angle $\psi_{\rm{B}} = 65.55^{\rm{o}}$ one observes $\Delta T \approx 0$ because vanishing surface reflectivity results in minimal interference effects. It is noteworthy that for the chosen thickness of the slab, which is $20$ times larger than the diameter of the reference sphere $d=1$ $\mu$m, the strength of the TFE is ten times smaller than in the sphere. Furthermore, in a $1$ $\mu$m-thick plate (not shown) the strength of the TFE plummets down because the interference effect is significantly weaker.

Of course, the fabrication techniques for magnetic films and plates are very much well-established and less effort demanding than those for single magnetic microspheres. However, the recent advances in making ultra-fine magnetic garnet particles make it possible to fabricate single scatterers with a close-to-spherical shape.\cite{38} YIG powders with controllable individual particle sizes of $\sim1$ $\mu$m can be obtained by a microwave heating method.\cite{39} Furthermore, a magnetic garnet microsphere can be fabricated using direct laser writing \cite{40}. Finally, laser-based printing techniques can potentially be applied.\cite{41}

\section{Conclusions}

We proposed an efficient scheme for enhancing the transverse Faraday effect in single magneto-dielectric microspheres. We investigated the scenario of a sphere made of a typical magnetic garnet exhibiting a realistically low absorption and high magneto-optical activity in the visible and infrared spectral ranges. We demonstrated the transverse Faraday effect of the order of $\sim10^{\rm{-3}}$, which is larger or comparable with typical values attainable with the modern MO devices which, however, have a significantly larger footprint as compared with microspheres. Our findings may find applications in nanophotonics, imaging, and magnonics.

\begin{acknowledgments}
This work was supported by the Australian Research Council. ISM gratefully acknowledges a postdoctoral research fellowship from the University of Western Australia. The authors thanks Ass./Prof. P. Metaxas for valuable discussions.
\end{acknowledgments}


\begin{thebibliography}{99}

\bibitem{1} M. Mansuripur, \textit{The Physical Principles of Magneto-Optical Recording} (Cambridge University Press, London, 1995).

\bibitem{2} A. K. Zvezdin and V. A. Kotov, \textit{Modern Magneto-Optics and Magneto-Optical Materials} (IOP Publishing, Bristol, 1997).

\bibitem{3} S. Sugano and N. Kojima, \textit{Magneto-Optics} (Springer, Berlin, 2000).

\bibitem{4} Ken-ichi Aoshima, N. Funabashi, K. Machida, Y. Miyamoto, K. Kuga, T. Ishibashi, N. Shimidzu, and F. Sato, J. Display Tech. \textbf{6,} 374 (2010).

\bibitem{5} V. V. Kruglyak, S. O. Demokritov, and D. Grundler, J. Phys. D: Appl. Phys. \textbf{43,} 264001 (2010).

\bibitem{6} A. A. Serga, A. V. Chumak, and B. Hillebrands, J. Phys. D: Appl. Phys. {\bf 43,} 264002 (2010).


\bibitem{7} Y. Souche, V. Novosad, B. Pannetier, and O. Geoffroy, J. Magn. Magn. Mater. {\bf 177--181,} 1277 (1998).

\bibitem{8} B. Bai, J. Tervo, and J. Turunen, New J. Phys. {\bf 8,} 205 (2006).

\bibitem{9} N. Kostylev, I. S. Maksymov, A. O. Adeyeye, S. Samarin, M. Kostylev, and J. F. Williams, Appl. Phys. Lett. {\bf 102,} 121907 (2013).

\bibitem{10} H. Marinchio, R. Carminati, A. Garc\'{i}a-Mart\'{i}n, and J. J. S\'{a}enz, New J. Phys. {\bf 16,} 015007 (2014).

\bibitem{11} I. S. Maksymov, J. Hutomo, and M. Kostylev, Opt. Express \textbf{22,} 8720 (2014).

\bibitem{12}  M. Inoue, M. Levy, and A. Baryshev, \textit{Magnetophotonics: From Theory to Applications} (Springer, Berlin, 2013).

\bibitem{13} J. Chen, P. Albella, Z. Pirzadeh, P. Alonso-Gonz\'{a}lez, F. Huth, S. Bonetti, V. Bonanni, J. \AA kerman, J. Nogu\'{e}s, P. Vavassori, A. Dmitriev, J. Aizpurua, and R. Hillenbrand, Small {\bf 7,} 2341 (2011).

\bibitem{14} R. Alcaraz de la Osa, J. M. Saiz, F. Moreno, P. Vavassori, and A. Berger Phys. Rev. B {\bf 85,} 064414 (2012).

\bibitem{15} G. Armelles, A. Cebollada, A. Garc\'{i}a-Mart\'{i}n, and M. U. Gonz\'{a}lez, Adv. Opt. Mater. {\bf 1,} 10 (2013).

\bibitem{16} N. de Sousa, L. S. Froufe-P\'{e}rez, G. Armelles, A. Cebollada, M. U. Gonz\'{a}lez, F. Garc\'{i}a, D. Meneses-Rodr\'{i}guez, and A. Garc\'{i}a-Mart\'{i}n, arXiv:1404.6767v1 [cond-mat.mes-hall] (2014).

\bibitem{17}  J. Y. Chin, T. Steinle, T. Wehlus, D. Dregely, T. Weiss, V. I. Belotelov, B. Stritzker, and H. Giessen, Nature Commun. \textbf{4,} 1599 (2013).

\bibitem{18} M. Mansuripur, Opt. Photon. News \textbf{10,} 32 (1999).

\bibitem{19} C. F. Bohren and D. R. Huffman, \textit{Absorption and Scattering of Light by Small Particles} (John Wiley and Sons, New York, 1983).

\bibitem{20} W. Hergert and Th. Wriedt, \textit{The Mie Theory: Basics and Applications} (Springer, Berlin, 2012).

\bibitem{21} M. Pohl, L. E. Kreilkamp, V. I. Belotelov, I. A. Akimov, A. N. Kalish, N. E. Khokhlov, V. J. Yallapragada, A. V. Gopal, M. Nur-E-Alam, and M. Vasiliev, New J. Phys. {\bf 15,} 075024 (2013).

\bibitem{22} D. Lacoste, B. A. van Tiggelen, G. L. J. A. Rikken, and A. Sparenberg, J. Opt. Soc. Am. A {\bf 15,} 1636 (1998).

\bibitem{23} R.-J. Tarento, K.-H. Bennemann, P. Joyes, and J. Van de Walle, Phys. Rev. E {\bf 69,} 026606 (2004).

\bibitem{23_1} E. D. Palik (Ed.), {\it Handbook of Optical Constants of Solids} (Academic Press, New York, 1985).

\bibitem{23_2} G. S. Krinchik and V. A. Artem'ev, Sov. Phys. JETP \textbf{26,} 1080 (1968).

\bibitem{24} A. E. Krasnok, I. S. Maksymov, A. I. Denisyuk, P. A. Belov, A. E. Miroshnichenko, C. R. Simovski and Y. S. Kivshar, Phys.-Usp. \textbf{56,} 539 (2013).

\bibitem{25} A. E. Krasnok, C. R. Simovski, P. A. Belov, and Y. S. Kivshar, ``Superdirective dielectric nanoantenna,'' Nanoscale accepted; doi:  10.1039/C4NR01231C.

\bibitem{26} R. Arias and D. L. Mills, Phys. Rev. B {\bf 70,} 104425 (2004).

\bibitem{27} P. Chu and D. L. Mills, Phys. Rev. B {\bf 75,} 054405 (2007).

\bibitem{28} H. T. Nguyen and M. G. Cottam, J. Appl. Phys. {\bf 103,} 07D513 (2008).

\bibitem{29} M. Bauer, C. Mathieu, S. O. Demokritov, B. Hillebrands, P. A. Kolodin, S. Sure, H. D\"{o}tsch, V. Grimalsky, Yu. Rapoport, and A. N. Slavin, Phys. Rev. B \textbf{56,} R8483(R) (1997).

\bibitem{30} F. Natali, B.J. Ruck, N.O.V. Plank, H.J. Trodahl, S. Granville, C. Meyer, and W. R. L. Lambrecht, Prog. Mater. Sci. \textbf{58,} 1316 (2013).

\bibitem{31} A. H. Lettington and C. Smith, \textit{Infra-red transparant materials} US Patent 5,723,207 (1998).

\bibitem{32} G. S. Abo, Y.-K. Hong, J. Park, J. Lee ,W. Lee, and B.-C. Choi, IEEE Trans. Magnet. \textbf{49,} 4937 (2013).

\bibitem{33} Z. Q. Qiu and S. D. Bader, Rev. Sci. Instrum. \textbf{80,} 043905 (2009).

\bibitem{34} D. A. Smith and K. L. Stockes, Opt. Express {\bf 14,} 5746 (2006). 

\bibitem{35} M. A. Yurkin, V. P. Maltsev, and A. G. Hoekstra, J. Quant. Spectr. Radiat. Transfer \textbf{106,} 546 (2007). 

\bibitem{35_1} J. Schneider and S. Hundson, IEEE Trans. Antenna Propagat. \textbf{41,} 944 (1993).

\bibitem{36} P. Yang and K. N. Liou, J. Opt. Soc. Am. A \textbf{13,} 2072 (1996).

\bibitem{37} S. Vi\v{s}\v{n}ovsk\'{y}, \textit{Optics in Magnetic Multilayers and Nanostructures} (CRC Press, Boca Raton, 2006).

\bibitem{38} G. C. Stangle, K. R. Venkatachari, S. P. Ostrander, W. A. Schulze, and J. D. Pietras, \textit{Process for making ultra-fine yttrium-iron-garnet particles} US Patent 5,660,773 (1997).

\bibitem{39} T. Kimura, H. Takizawa, K. Uheda, T. Endo, and M. Shimada, J. Am. Ceram. Soc. \textbf{81,} 2961 (1998).

\bibitem{40} T. Amemiya, A. Ishikawa, Y. Shoji, Pham Nam Hai, M. Tanaka, T. Mizumoto, T. Tanaka, and S. Arai, Opt. Lett. {\bf 39,} 212 (2014).

\bibitem{41} C. L. Sones, M. Feinaeugle, A. Sposito, B. Gholipour, and R. W. Eason, Opt. Express \textbf{20,} 15171 (2012).



\end{thebibliography}
\end{document}